\documentstyle[preprint,eqsecnum,aps]{revtex}
\begin{document}
\draft
\title{Can the Wigner function be determined by properties for translation and parity
transformation on lattice phase space?}
\author{Minoru HORIBE, Takaaki HASHIMOTO and Akihisa HAYASHI}
\address{
Department of Applied Physics\\
Fukui University, Fukui 910, Japan
}
\date{\today}
\maketitle
\begin{abstract}
We show that the Fano operator for one dimensional quantum system is uniquely
determined by assuming the reasonable behavior under translation and parity transformation on phase space.  
Contrarily, for the system with lattice phase space the same procedure does not work. 
\end{abstract}
\section{Introduction}
The expectation values for mixed state with density matrix $\hat{\rho}$ are expressed
as the averages over phase space quasiprobability $W(q,p)$ defined by
\begin{equation}
W(q,p)=\frac{1}{2 \pi \hbar}\int_{-\infty}^{\infty}dr
  \left[
      e^{-ipr/\hbar} 
      \left\langle q+\frac{r}{2} \right| \hat{\rho} 
      \left|       q-\frac{r}{2} \right\rangle
   \right],
\label{wignerf}
\end{equation}
which is well known as the Wigner function\cite{wigner}, where
$\left|q \pm \frac{r}{2} \right\rangle$ is eigenvector with eigenvalue 
$q \pm \frac{r}{2}$ for the coordinate operator.
We can check that this function satisfies the following conditions

\vspace{2mm}

\noindent
(A) We can obtain the marginal distribution along the 
coordinate and momentum axes,
\begin{eqnarray*}
\int^{\infty}_{-\infty}W(q,p)dp&=&\langle q |\hat{\rho}|q \rangle, \\
\int^{\infty}_{-\infty}W(q,p)dq&=&\langle p |\hat{\rho}|p \rangle.
\end{eqnarray*} 
(B) The Wigner function is a real valued function,
\[
W^{\ast}(q,p)=W(q,p).
\]
(C) The Wigner function includes the same information as the density
matrix.\\

\vspace{2mm}

\noindent Here, $|q \rangle$ and $|p \rangle$ are eigenvectors for coordinate operator
$\hat{Q}$ and momentum operator $\hat{P}$, respectively,
\[
\hat{Q}|q \rangle =q|q \rangle,\;\;\;\hat{P}|p \rangle =p| p \rangle
\]

Inversely, it is pointed out that the Wigner function is not determined uniquely under these conditions by many physicists \cite{wigner2} $\sim$ \cite{hashimoto}.
From tomographic point of view, Bertrand and Bertrand \cite{BertTomo} and 
Leonhardt\cite{LeonQO} impose an additional condition which gives the connection between rotations
of quantum variables $(\hat{Q},\hat{P})$ and of the point $(q,p)$ in phase space on which the Wigner function is defined, 
\[
\langle q|R_{\theta}\hat{\rho}R^{-1}_{\theta}|q \rangle=
\int^{\infty}_{\infty}W(q\cos\theta+p\sin\theta,-q\sin\theta+p\sin\theta)dp,
\]
where $R_{\theta}$ is unitary operator for rotation of quantum variables $(\hat{Q},\hat{P})$,
\[
R_{\theta}\hat{Q}R^{-1}_{\theta}=\hat{Q}\cos\theta-\hat{P}\sin\theta,\;\;
R_{\theta}\hat{P}R^{-1}_{\theta}=\hat{P}\cos\theta+\hat{Q}\sin\theta.
\]
And they show that there is only one
solution satisfying it for one dimensional quantum system. 

In the previous paper
\cite{horibe}, we rewrote this condition using the Fano operator $\hat{\Delta}(q,p)$\cite{fano} defined by
\begin{eqnarray}
    W(q,p)&=&{\rm Tr}[\hat{\Delta}(q,p)\hat{\rho}],\label{codeffa} \\
\hat{\rho}&=&\frac{1}{2\pi\hbar}\int^{\infty}_{\infty} W(q,p)\hat{\Delta}^{\dag}(q,p)dqdp. 
\label{coideffa}
\end{eqnarray}
Namely, we assumed that
\begin{equation}
R_{\theta}\hat{\Delta}(q,p)R^{-1}_{\theta}=\hat{\Delta}(q\cos\theta+p\sin\theta,-q\sin\theta+p\sin\theta),
\label{controt}
\end{equation}
and showed that there is only one solution satisfying this condition. For the lattice phase space with
$N^2$ sites, we could discuss it in the same manner and we found the unique solution,
which is equivalent to the one given by Cohendet\cite{cohendet}, for the case where 
$N$ is odd, but no solution for the case where $N$ is even.
 Naively, we are interested in whether the Wigner function is determined uniquely under the assumption for the properties of simpler transformation than rotation. In this paper, we assume the behavior of the Fano operator$\hat{\Delta}(q,p)$ in equation(\ref{codeffa}) under the translation and parity transformation
and try to determine the Fano operator.
For one dimensional quantum system, we can find out only one Fano operator which satisfies the conditions corresponding to the above ones (A) $\sim$ (C) and new conditions. 
But, for systems with lattice phase space, we cannot determine it uniquely.

\section{the Wigner function on one dimensional system}
In this section, we study the Fano operator $\hat{\Delta}(q,p)$ defined by equations (\ref{codeffa}) and
(\ref{coideffa}). In terms of the Fano operator,
we can rewrite the conditions (A) $\sim$ (C) in the preceding section,
\begin{eqnarray}
\int^{\infty}_{-\infty}\hat{\Delta}(q,p)dp&=&|q \rangle\langle q|, \label{condm1}\\
\int^{\infty}_{-\infty}\hat{\Delta}(q,p)dq&=&|p \rangle\langle p|, \label{condm2}\\
\hat{\Delta}^{\dag}(q,p)&=&\hat{\Delta}(q,p),\label{condm3}\\
\int^{\infty}_{-\infty}\int^{\infty}_{-\infty}
(\hat{\Delta}^{\dag})_{q_1q_2}(q,p)(\hat{\Delta})_{q_3q_4}(q,p)dpdq&=&
\frac{1}{2\pi\hbar}\delta(q_1-q_4)\delta(q_3-q_2).\label{condm4}
\end{eqnarray}
where $|q\rangle$ and $|p \rangle$ are eigenvectors of the coordinate operator $\hat{Q}$ and the momentum operator $\hat{P}$ with eigenvalues $q$ and $p$, respectively,
as stated previously and $\Delta_{q_1q_2}(q,p)=\langle q_1|\hat{\Delta}(q,p) |q_2 \rangle$ is a matrix element of the Fano
operator between eigenvectors of the operator $\hat{Q}$ for eigenvalues $q_1$ and $q_2$.  
When we expand the Fano operator in terms of complete set
$e^{i{\cal Q}\hat{P}/\hbar}e^{-i{\cal P}\hat{Q}/\hbar}$,
\begin{equation}
\hat{\Delta}(q,p)=\frac{1}{2\pi\hbar}\int^{\infty}_{-\infty}d{\cal Q}d{\cal P}
a(q,p;{\cal Q},{\cal P})e^{i{\cal Q}\hat{P}/\hbar}e^{-i{\cal P}\hat{Q}/\hbar},
\label{expfano}
\end{equation}
the coefficients $a(q,p;{\cal Q},{\cal P})$ should satisfy the conditions
\begin{eqnarray*}
\int^{\infty}_{-\infty}a(q,p;{\cal Q},{\cal P})dp&=&
\delta({\cal Q})e^{iq{\cal P}/\hbar},\\
\int^{\infty}_{-\infty}a(q,p;{\cal Q},{\cal P})dq&=&
\delta({\cal P})e^{-ip{\cal Q}/\hbar},\\
a^{\ast}(q,p;{\cal Q},{\cal P})&=&
e^{-i{\cal Q}{\cal P}/\hbar}a(q,p;{\cal Q},{\cal P}),\\
\int^{\infty}_{\infty}a^{\ast}(q,p:{\cal Q},{\cal P})
a(q,p:{\cal Q}',{\cal P}')dqdp&=&
\delta({\cal Q}-{\cal Q}')\delta({\cal P}-{\cal P}'),
\end{eqnarray*}
because of conditions (\ref{condm1}) $\sim$ (\ref{condm4}).
Using the Fourier transformed coefficients $\tilde{a}(s,t;{\cal Q},{\cal P})$
\begin{equation}
\tilde{a}(s,t;{\cal Q},{\cal P})=
\frac{1}{2\pi\hbar}\int^{\infty}_{-\infty}a(q,p;{\cal Q},{\cal P})e^{-i(qs-pt)/\hbar}dqdp,
\end{equation}
these conditions can be described in simpler forms;
\begin{eqnarray}
&&\tilde{a}(s,0;{\cal Q},{\cal P})=\delta({\cal Q})\delta({\cal P}-s),
\label{cofcon1}\\
&&\tilde{a}(0,t;{\cal Q},{\cal P})=\delta({\cal Q}-t)\delta({\cal P}),
\label{cofcon2}\\
&&\tilde{a}(s,t;{\cal Q},{\cal P})^{\ast}=\tilde{a}(-s,-t;-{\cal Q},-{\cal P})
e^{-i{\cal Q}{\cal P}/\hbar},
\label{cofcon3}\\
&&
\int^{\infty}_{-\infty}dsdt
\tilde{a}(s,t;{\cal Q}',{\cal P}')^{\ast}\tilde{a}(s,t;{\cal Q},{\cal P}) 
=\delta({\cal Q}-{\cal Q}')\delta({\cal P}-{\cal P}').
\label{cofcon4}
\end{eqnarray} 

\subsection{New conditions arising from translation and parity transformation}
For the classical theory, the distribution function $\rho(q,p)$ on phase space is transformed by the following way;
\[
\rho'(q',p')=\rho(q,p)=\rho(q'-a,p'-b),
\]
under the translation on phase space,
\[
q \rightarrow q'=q+a,\;\; p \rightarrow p'=p+b.
\]
And under the parity transformation on phase space 
\[
q \rightarrow q'=-q,\;\; p \rightarrow p'=-p,
\]
the distribution function is changed into $\rho'(q',p')$
\[
\rho'(q',p')=\rho(q,p)=\rho(-q',-p').
\]
Thus, in the quantum theory,  we hope that the Fano operator is transformed as follows;
\begin{equation}
U_{{\rm cont}}(a,b)\hat{\Delta}_(q,p)U_{{\rm cont}}^{-1}(a,b)=\Delta(q-a,p-b),
\label{transfano}
\end{equation}
and 
\begin{equation}
T_{{\rm cont}}\hat{\Delta}(q,p)T_{{\rm cont}}^{-1}=\hat{\Delta}(-q,-p).\label{parityfano}
\end{equation}
Here, $U_{{\rm cont}}(a,b)$ and $T_{{\rm cont}}$ are unitary operators which are defined by
\begin{eqnarray}
U_{{\rm cont}}(a,b)&=&\exp\left[i\frac{\hat{P}a-\hat{Q}b}{\hbar}\right], 
\label{transfunit}\\
T_{{\rm cont}}&=&\exp\left[-i\pi\frac{\hat{Q}^2+\hat{P}^2}{2\hbar}\right].
\label{parityunit}
\end{eqnarray}
It is easily shown that these unitary operators $U_{{\rm cont}}(a,b)$ and $T_{{\rm cont}}$ induce the translation and parity transformation, respectively,
\begin{eqnarray}
  U_{{\rm cont}}(a,b)\hat{Q}U_{{\rm cont}}^{-1}(a,b)=\hat{Q}+a,\;\;\;
&&U_{{\rm cont}}(a,b)\hat{P}U_{{\rm cont}}^{-1}(a,b)=\hat{P}+b\\
\label{transfvarQ}
  T_{{\rm cont}}\hat{Q}T_{{\rm cont}}^{-1}=-\hat{Q},\;\;\;
&&T_{{\rm cont}}\hat{P}T_{{\rm cont}}^{-1}=-\hat{P}
\label{transfvarP}
\end{eqnarray}
For the coefficients in the expansion (\ref{expfano}), these conditions (\ref{transfano}) and (\ref{parityfano}) become
\begin{eqnarray}
a(q,p;{\cal Q},{\cal P})e^{i{\cal Q}b/\hbar}e^{-ia{\cal P}/\hbar}
&=&a(q-a,p-b;{\cal Q},{\cal P}),\label{transcoff}\\
a(q,p;-{\cal Q},-{\cal P})&=&a(-q,-p;{\cal Q}{\cal P}).
\label{paritycoff}
\end{eqnarray}
Using the Fourier transformed coefficients $\tilde{a}(s,t;{\cal Q},{\cal P})$,
these conditions are given by
\begin{eqnarray}
\tilde{a}(s,t;{\cal Q},{\cal P})&=&
e^{-i({\cal Q}-t)b/\hbar}e^{ia({\cal P}-s)/\hbar}
\tilde{a}(s,t;{\cal Q},{\cal P})
\label{transcofur}\\
\tilde{a}(s,t;-{\cal Q},-{\cal P})&=&
\tilde{a}(-s,-t;{\cal Q},{\cal P})
\label{paritycofur}
\end{eqnarray}
Thus, we have simple equations from the assumptions (\ref{transfano}) and
(\ref{parityfano}). In the next subsection we try to find the Fano operator
satisfying conditions (\ref{cofcon1}) $\sim$ (\ref{cofcon4}), (\ref{transcofur})
and (\ref{paritycofur}). 

\subsection{the Fano operator under the new conditions}
From the equation (\ref{transcofur}), we have
\begin{equation}
\tilde{a}(s,t;{\cal Q},{\cal P})=F(s,t)\delta({\cal Q}-t)\delta({\cal P}-s)
\label{newcon1}
\end{equation}
where $F(s,t)$ is a function of $s$ and $t$ which is determined by other conditions.

Taking account of the condition (\ref{paritycofur}), we obtain the condition for 
the function $F(s,t)$
\begin{equation}
F(s,t)=F(-s,-t).
\label{newcon2}
\end{equation}
From conditions (\ref{newcon2}) and (\ref{cofcon3}), the function $F(s,t)$
should  satisfy
\[
F^{\ast}(s,t)=e^{-its/\hbar}F(-s,-t)=e^{-its/\hbar}F(s,t).
\]
So that we get
\[
F(s,t)=R(s,t)e^{ist/2\hbar},
\]
where $R(s,t)$ is a real function of $s$ and $t$. The value of the square of this function
$R(s,t)$ is restricted to unity by the condition (\ref{cofcon4})
\[
R^2(s,t)=1\;\;{\rm or}\;\;R(s,t)=\pm 1.
\]
Because of the condition (\ref{cofcon1}), we should choose $+1$ as $R(s,t)$ and we have
unique solution
\[
\tilde{a}(s,t;{\cal Q},{\cal P})=e^{-i{\cal Q}{\cal P}/\hbar}
\delta({\cal Q}-t)\delta({\cal P}-s).
\]
Here we assumed that the function $R(s,t)$ is continuous.
\section{the Wigner function on lattice phase space}
In this section, we try to determine the Fano operator by the similar method to the one we adopted in the preceding section.
For lattice phase space with $N^2$ sites,
we can obtain the conditions corresponding to the conditions (A) $\sim$ (C)  
by replacing the integration by summation over $Z_N$ in the equations (\ref{cofcon1}),
(\ref{cofcon2}), (\ref{cofcon3}) and (\ref{cofcon4}),
\begin{eqnarray}
\sum_{p \in Z_N}\hat{\Delta}(q,p)&=& |q \rangle \langle q |, \label{discon1} \\
\sum_{q \in Z_N}\hat{\Delta}(q,p)&=& |p \rangle \langle p |, \label{discon2} \\
\hat{\Delta}^{\dag}(q,p)&=&\hat{\Delta}(q,p),   \label{discon3} \\
\sum_{q,p \in Z_N}(\hat{\Delta}^{\dag})_{q_1q_2}(q,p)(\hat{\Delta})_{q_3q_4}(q,p)&=&
\frac{1}{N}\delta^{(N)}_{q_1q_4}\delta^{(N)}_{q_3q_2}.
\label{discon4}
\end{eqnarray}
where $|q\rangle$ and $|p \rangle$ are eigenvectors of ``coordinate" and ``momentum" 
operators with eigenvalue $q\;(q \in Z_N)$ and $p\;(p \in Z_N)$, respectively and $\delta^{(N)}_{q_1q_2}$ is
Kronecker's delta on $Z_N$,
\[
\delta^{(N)}_{q_1q_2}=\left\{\begin{array}{cc}
  1 &  (q_1\;\;{\rm mod}\;\;N=q_2) \\
  0 &  (q_1\;\;{\rm mod}\;\;N\neq q_2)
  \end{array}. \right.
\]

We decompose the Fano operator $\hat{\Delta}(q,p)$ into matrices $S^nP^m\;\;(n,m=0,1,2,\cdots, N-1),$
\begin{equation}
\hat{\Delta}(q,p)=\sum_{n,m \in Z_N}a(q,p;n,m)S^nP^m
\label{laFanoex}
\end{equation}
where the matrices $S$ and $P$ are defined by
\begin{eqnarray}
S&=&\left(\begin{array}{cccc}
              0   & 1      & \cdots &   0    \\
           \vdots & \ddots & \ddots & \vdots \\
           \vdots &        & \ddots &   1    \\
              1   &        &        &   0
        \end{array} \right) \label{shift} \\
 P&=&\left(\begin{array}{ccccc}
              1   & 0      & \cdots   & \cdots &   0    \\
              0   & \ddots &          &        & \vdots \\
           \vdots &        & \omega^n &        & \vdots \\
           \vdots &        &          & \ddots &   0    \\
              0   &        &          &   0    &  \omega^{(N-1)}
        \end{array} \right) \label{phase}
\end{eqnarray}
and $\omega$ is a primitive N-th root of unity,
\[
\omega=e^{\frac{2\pi i}{N}}.
\]
These matrices satisfy commutation relation
\begin{equation}
S^nP^m=\omega^{nm} P^mS^n, (n,m={\rm integer})
\label{commSP}
\end{equation}
This commutation relation appears similar to the commutation relation between operators $e^{-i{\cal P}\hat{Q}/\hbar}$ and $e^{ i{\cal Q}\hat{P}/\hbar}$ used in the preceding section
\[
 e^{ i{\cal Q}\hat{P}/\hbar}e^{-i{\cal P}\hat{Q}/\hbar}
=e^{-i{\cal P}{\cal Q}/\hbar}e^{-i{\cal P}\hat{Q}/\hbar}e^{ i{\cal Q}\hat{P}/\hbar},
\]
and we can think that matrices $P$ and $S$ correspond to  
$e^{-i\hat{Q}/\hbar}$ and $e^{ i\hat{P}/\hbar}$, respectively. 
It is natural that the eigenvector $|q \rangle$ and $|p \rangle$ are regarded as
the eigenvectors with respect to the matrices $P$ and $S$, respectively:
\begin{eqnarray}
P|q \rangle &=& \omega^q    |q \rangle,  \label{eigenP}\\
S|p \rangle &=& \omega^{-p} |p \rangle.  \label{eigenS}
\end{eqnarray}
So, the eigenvector $|p \rangle$ is expressed as the linear combination of the eigenvector
$|q \rangle$,
\[
|p \rangle=\frac{1}{\sqrt{N}}\sum_{Z_N \in q}\omega^{-pq}|q \rangle.
\]
We will use this correspondence of $P$ and $S$ to $e^{-i\hat{Q}/\hbar}$
and $e^{ i\hat{P}/\hbar}$
in order to define the translation and parity transformation in lattice phase space.  

Using the coefficients $a(q,p;n,m)$, the conditions (\ref{discon1}) $\sim$ (\ref{discon4}) become
\begin{eqnarray}
&&\sum_{p \in Z_N}a(q,p;n,m)=\frac{1}{N}\omega^{-qm}\delta_{n,0},
\label{lcon1}\\
&&\sum_{q \in Z_N}a(q,p;n,m)=\frac{1}{N}\omega^{pn}\delta_{m,0},
\label{lcon2}\\
&&a(q,p;n,m)=\omega^{-nm}a(q,p:N-n,N-m),
\label{lcon3}\\
&&\sum_{q,p \in Z_N}a(q,p;n,m)a(q,p;k,l)=\frac{1}{N^2}\delta^{(N)}_{m,k}\delta^{(N)}_{n,l},
\label{lcon4}
\end{eqnarray}
In order to obtain the equation (\ref{lcon4}), we used the relation,
\[
\delta^{(N)}_{a,a'}\delta^{(N)}_{b,b'}=\frac{1}{N^2}\sum_{n,m \in Z_N}
(S^nP^m)_{ab}(S^nP^{-m})_{a'b'}.
\]
We introduce the Fourier transformed coefficients $\tilde{a}(s,t;n,m)$
\begin{equation}
\tilde{a}(s,t;n,m)=\frac{1}{N^2}\sum_{q,p \in Z_N}\omega^{qs}\omega^{-pt}
a(q,p;n,m).
\label{lfourco}
\end{equation}
The above conditions (\ref{lcon1}) $\sim$ (\ref{lcon4}) become
\begin{eqnarray}
&&\tilde{a}(s,0;n,m)=\frac{1}{N^2}\delta^{(N)}_{n,0}\delta^{(N)}_{m,s},
\label{lcon1f} \\
&&\tilde{a}(0,t;n,m)=\frac{1}{N^2}\delta^{(N)}_{n,t}\delta^{(N)}_{m,0},
\label{lcon2f} \\
&&\tilde{a}(s,t;n,m)=\omega^{-nm}\tilde{a}^{\ast}(-s,-t;-n,-m),
\label{lcpn3f} \\
&&\sum_{s,t \in Z_N}\tilde{a}(s,t;n,m)^{\ast}\tilde{a}(s,t;k,l)=
\frac{1}{N^4}\delta^{(N)}_{n,k}\delta^{(N)}_{m,l}.
\label{lcon4f}
\end{eqnarray}

\subsection{New conditions arising from translation and parity transformation}
As we explained in the preceding subsection, since we consider that the matrices
$P$ and $S$ as $e^{-i\hat{Q}/\hbar}$  and  $e^{i\hat{P}/\hbar}$, 
from the definition (\ref{transfunit}) of $U_{{\rm cont}}(a,b)$, 
the unitary matrices $U_{{\rm dis}}(a,b)$ for the translation in lattice
phase space is obtained,
\begin{equation}
U_{{\rm cont}}(a,b)=\exp\left[i\frac{\hat{P}a-\hat{Q}b}{\hbar}\right]
\rightarrow  U_{{\rm dis}}(a,b)=P^bS^a,
\label{disdefU}
\end{equation}
From commutation relation (\ref{commSP}), we obtain
\begin{equation}
\left\{\begin{array}{c}
P \rightarrow P'=U_{{\rm dis}}(a,b)PU_{{\rm dis}}^{-1}(a,b)=\omega^{ a}P, \\
S \rightarrow S'=U_{{\rm dis}}(a,b)SU_{{\rm dis}}^{-1}(a,b)=\omega^{-b}S, 
\end{array}\right. 
\label{dtrans}\\
\end{equation}
This transformation is similar to the transformation of $e^{-i\hat{Q}/\hbar}$
and $e^{i\hat{P}/\hbar}$ by unitary operator $U_{{\rm cont}}(a,b)$.
 
Imitating the equation (\ref{transfano}), we assume that the Fano operator satisfies 
\begin{equation}
U_{{\rm dis}}(a,b)\hat{\Delta}_(q,p)U_{{\rm dis}}^{-1}(a,b)=\Delta(q-a,p-b).
\label{distrafa}
\end{equation}
Similarly, for the parity transformation, we assume as follows
\begin{equation}
T_{{\rm dis}}\hat{\Delta}(q,p)T^{-1}_{{\rm dis}}=\hat{\Delta}(-q,-p),
\label{disparfa}
\end{equation}
where $T_{{\rm dis}}$ is the unitary matrix for the parity transformation, which is given by
\begin{eqnarray} 
T_{\rm dis}&=&\left(\begin{array}{cccccc}
         1     &    0   &    0   & \cdots &   0    &   0    \\
         0     &    0   &    0   & \cdots &   0    &   1    \\
         0     &    0   &    0   & \cdots &   1    &   0    \\
        \vdots & \vdots & \vdots &        & \vdots & \vdots \\
         0     &    0   &    1   & \cdots &    0   &   0    \\
         0     &    1   &    0   & \cdots &    0   &   0     
         \end{array}\right) \nonumber\\
{\rm or}\;\;
(T)_{\alpha, \beta}&=&\delta^{(N)}_{\alpha,N-\beta}. \label{disdefT}
\end{eqnarray}
It is checked that this unitary matrix $T_{{\rm dis}}$ transforms matrices $P$ and $S$ into
inverse matrices $P^{-1}$ and $S^{-1}$, respectively,
\begin{equation}
\left\{\begin{array}{c}
P \rightarrow P'=T_{{\rm dis}}PT_{{\rm dis}}^{-1}=P^{-1}, \\
S \rightarrow S'=T_{{\rm dis}}ST_{{\rm dis}}^{-1}=S^{-1},
\end{array}\right. \label{dparit}
\end{equation}

For the coefficients $a(q,p;n,m)$ in the expansion (\ref{laFanoex}),
these conditions (\ref{distrafa}) and (\ref{disparfa}) become,

\begin{eqnarray*}
a(q,p;n,m)&=&\omega^{-ma+nb}a(q-a,p-b;n,m), \\
a(q,p;n,m)&=&a(-q,-p;-n,-m),                
\end{eqnarray*}
and the coefficients $\tilde{a}(s,t;n,m)$ in equation (\ref{lfourco}) satisfy 
\begin{eqnarray}
\tilde{a}(s,t;n,m)&=&\omega^{a(s-m)}\omega^{-b(t-n)}\tilde{a}(s,t;n,m),
\label{distracof}\\
\tilde{a}(s,t;n,m)&=&\tilde{a}(-s,-t;-n,-m).
\label{disparcof}
\end{eqnarray}
                   
\subsection{the Fano operator under the new conditions}
In order that the coefficients $\tilde{a}(s,t;n,m)$ satisfy the condition (\ref{distracof})
for arbitrary integers $a$ and $b$, $\tilde{a}(s,t;n,m)$ should be proportional to
$\delta^{(N)}_{t,n}\delta^{(N)}_{s,m}$,
\begin{equation}
\tilde{a}(s,t;n,m)=F(s,t)\delta^{(N)}_{t,n}\delta^{(N)}_{s,m},
\label{dinewcon1}
\end{equation}
where $F(s,t)$ is a complex valued function. And, from the condition(\ref{disparcof}), the function $F(s,t)$ should be a
symmetric function under the reflection $ s \rightarrow -s,\; t\rightarrow -t$
\begin{equation}
F(-s,-t)=F(s,t).
\label{disconF}
\end{equation}
The condition (\ref{lcpn3f}) with the above equation determines the phase factor
up to sign,
\begin{equation}
F(s,t)=\omega^{-st/2}R(s,t),
\label{disconF1}
\end{equation}
where $R(s,t)$ is a real function.
Substituting this equation (\ref{disconF1}) into the condition (\ref{lcon4f}),
we can see that
\[
R(s,t)^2=1.
\]
Thus, we obtain the same condition as we did for one dimensional quantum system in the
preceding section. In that case, we could determine the $R(s,t)$ since we assumed that it is
a continuous
function. However, we have many ways of assigning $\pm1$ to each site on a lattice phase
space. For example, in the case where $N$ is odd, if we choose
\[
R(s,t)=(-1)^{st}=\omega^{Nst/2},
\]
we get the Fano operator which is given by Cohendet et al.\cite{cohendet}.  

\section{Summary and Discussion}
We tried to determine the Fano operator uniquely under the assumptions for translation
and parity transformation. For one dimensional quantum system, we found out only one Fano operator satisfying these conditions and three original conditions. 
Contrary to this case, for the lattice phase space which includes $N^2$ sites,
we could not determine the Fano operator uniquely.

We considered the map from a point on phase space to the point rotated
about the original point $(q,p)=(0,0)$ by $\pi$ as the parity transformation.
However, there are quantum systems where we had better consider the rotation
about another point, instead of the origin, as the parity transformation. For example, in spin
systems, the parity transformation is corresponding to exchanging between eigenstates for
eigenvalues $\omega^{k}$ and $\omega^{N-1-k}$ of matrix $S$. This transformation is equivalent
to the rotation about the point $\left(\frac{N-1}{2},\frac{N-1}{2}\right)$. 
The rotation $T_{{\rm cont}}(c,d)$ and $T_{{\rm dis}}(c,d)$ about the point $(c/2,d/2)$ by $\pi$ for continuous and discrete phase space can be described in terms of combination of translation and 
the rotation about the origin;
\begin{equation}
(q,p) 
\stackrel{U^{-1}\left(\frac{c}{2},\frac{d}{2}\right)}{\longrightarrow}
\left(q-\frac{c}{2},p-\frac{d}{2} \right)
\stackrel{T}{\longrightarrow}
\left(\frac{c}{2}-q,\frac{d}{2}-p \right)
\stackrel{U\left(\frac{c}{2},\frac{d}{2}\right)}{\longrightarrow}
(c-q,d-p),
\end{equation}
and we have 
\begin{equation}
   T(c,d)=U(c,d)T. \label{paranp}
\end{equation}
Hereafter, the subscripts ``cont" and ``dis" are dropped, as we have the same equations for both continuous and discrete phase spaces.
From the equation (\ref{paranp}), we can expect that the assumptions for this transformation do not give rise to essentially different conditions from the ones we considered in the preceding sections. 
Indeed, if we assume the behaviors
\begin{equation}
T(c,d)\hat{\Delta}(q,p)T^{-1}(c,d)=\hat{\Delta}(c-q,d-p),  
\end{equation}
we have 
\begin{eqnarray}
\tilde{a}(s,t;{\cal Q},{\cal P})&=&e^{-id({\cal Q}-t)/\hbar}e^{ic({\cal P}-s)/\hbar}
\tilde{a}(-s,-t;{\cal Q},{\cal P}), \label{newcopac}\\
\tilde{a}(s,t;n,m)&=&\omega^{c(s-m)}\omega^{-d(t-n)}
\tilde{a}(-s,-t;n,m). \label{newcopad}
\end{eqnarray}
Owing to the delta functions or Kronecker's delta in equation(\ref{newcon1})
and (\ref{dinewcon1}), the factors $e^{-id({\cal Q}-t)/\hbar}e^{ic({\cal P}-s)/\hbar}$ 
and $\omega^{c(s-m)}\omega^{-d(t-n)}$ are vanished, so that these conditions reduce to the ones
(\ref{paritycofur}) and (\ref{disparcof}) we considered. 


\end{document}